\documentclass[aps,prl,twocolumn,
			   groupedaddress,superscriptaddress,
			   amsfonts,amssymb,amsmath,
			   citeautoscript,longbibliography,
			   a4paper
			   ]{revtex4-2}

\usepackage[utf8]{inputenc}
\usepackage[english]{babel}

\usepackage{microtype} %For better kerning and symbol-stretching
\usepackage{xspace} %For the \xspace command

\usepackage{txfonts}  %Times Roman fonts
\usepackage{txfontsb} %Addition for txfonts, including old style numerals and greek

\usepackage{bm} %Bold math symbols with \bm{} (Greek and other symbols)

\usepackage{xcolor}
\usepackage[]{graphicx} % "demo" option to disable rendering for speed
\graphicspath{{figs/}}

\usepackage[]{booktabs}
\usepackage{array}
\usepackage{layouts}
\usepackage{multirow}

\usepackage{enumerate}
\usepackage[inline]{enumitem}

\usepackage{algorithm}
\usepackage{algpseudocode}

\usepackage{xr}
\makeatletter
\newcommand*{\addFileDependency}[1]{% argument=file name and extension
  \typeout{(#1)}% latexmk will find this if $recorder=0 (however, in that case, it will ignore #1 if it is a .aux or .pdf file etc and it exists! if it doesn't exist, it will appear in the list of dependents regardless)
  \@addtofilelist{#1}% if you want it to appear in \listfiles, not really necessary and latexmk doesn't use this
  \IfFileExists{#1}{}{\typeout{No file #1.}}% latexmk will find this message if #1 doesn't exist (yet)
}
\makeatother

\usepackage{hyperref}
\hypersetup{colorlinks,
	linkcolor={blue!75!black!80!yellow},
	citecolor={blue!75!black!80!yellow},
	urlcolor={blue!75!black!80!yellow}
}

\usepackage[capitalize,nameinlink]{cleveref}

\crefname{subequations}{Eqs.}{Eqs.} %Specific changes to allow for Eqs.-wording when referring to a set of subequations. Label of subequations must include [subequations] as an option.
\Crefname{subequations}{Eqs.}{Eqs.}
\crefformat{subequations}{#2Eqs.~(#1)#3}
\Crefformat{subequations}{#2Eqs.~(#1)#3}
\crefname{page}{p.}{p.} %Changing from 'page' to 'p.'

\usepackage{placeins}

\usepackage{siunitx}
\DeclareSIUnit[number-unit-product = ]\percent{\char`\%} % remove spacing for \percent

\usepackage[centering,hmargin=16mm,tmargin=30mm,bmargin=26mm]{geometry}

\usepackage{soul}

\usepackage{textcomp} % for \textrightarrow
\usepackage{xifthen}
\usepackage{etoolbox}
\newboolean{togglecomments}
\newboolean{toggletodos}
\newboolean{togglechanges}

\setboolean{togglecomments}{true}
\setboolean{toggletodos}{true}
\setboolean{togglechanges}{false}

\newcommand{\rv}{\mathbf{r}}

\newcommand{\kv}{\mathbf{k}}
\newcommand{\Ev}{\mathbf{E}}
\newcommand{\Dv}{\mathbf{D}}

\newcommand{\Rv}{\mathbf{R}}
\newcommand{\Gv}{\mathbf{G}}

\newcommand{\nv}{\mathbf{n}}

\newcommand{\e}{\mathrm{e}}

\newcommand{\iu}{\mathrm{i}}

\newcommand{\ie}{i.e.,\@\xspace} %Gobble-spaces of the "small" type (small is ensured by adding \@)

\newcommand{\eg}{e.g.,\@\xspace}

\newcommand{\appropto}{\mathrel{\vcenter{
			\offinterlineskip\halign{\hfil$##$\cr
				\propto\cr\noalign{\kern.2pt}\sim\cr\noalign{\kern-2.5pt}}}}}

\let\Re\relax %Remove the default definition before redefining
\DeclareMathOperator{\Re}{Re}
\DeclareFontFamily{U}{mathx}{\hyphenchar\font45}
\DeclareFontShape{U}{mathx}{m}{n}{<5> <6> <7> <8> <9> <10>
                                  <10.95> <12> <14.4> <17.28> <20.74> <24.88>
                                  mathx10}{}
\DeclareSymbolFont{mathx}{U}{mathx}{m}{n}
\DeclareFontSubstitution{U}{mathx}{m}{n}
\makeatletter
\newcommand{\raisemath}[1]{\mathpalette{\raisem@th{#1}}}
\newcommand{\raisem@th}[3]{\raisebox{#1}{$#2#3$}}
\makeatother

\renewcommand{\paragraph}[1]{\vskip 1ex\noindent\textbf{#1.}~}

\usepackage[eulergreek]{sansmath}
\makeatletter
\renewcommand\@make@capt@title[2]{%
    \@ifx@empty\float@link{\@firstofone}{\expandafter\href\expandafter{\float@link}}%
    \sansmath\sffamily\textbf{#1\@caption@fignum@sep}#2 % does not work with the newtx* packages unfortunately
}%

\makeatother

\graphicspath{{figures/}}

\setboolean{togglecomments}{true}
\setboolean{toggletodos}{true}
\setboolean{togglechanges}{false}
\begin{document}

%------------------------------------
%----- AUTHORS AND AFFILIATIONS -----
%------------------------------------
\newcommand{\miteeaffil}{Department of Electrical Engineering and Computer Science, Massachusetts Institute of Technology, Cambridge, Massachusetts, USA}
\newcommand{\mitphysaffil}{Department of Physics, Massachusetts Institute of Technology, Cambridge, Massachusetts, USA}
\newcommand{\mitmathaffil}{Department of Mathematics, Massachusetts Institute of Technology, Cambridge, Massachusetts, USA}

\author{Samuel Kim}
\email{samkim@mit.edu}
\affiliation{\miteeaffil}
\author{Thomas Christensen}
\email{tchr@mit.edu}
\affiliation{\mitphysaffil}
\author{Steven G.
Johnson}
\affiliation{\mitphysaffil}
\affiliation{\mitmathaffil}
\author{Marin Solja\v{c}i\'{c} }
\affiliation{\mitphysaffil}

%---------------------------
%----- KEYWORDS & PACS -----
%---------------------------

\title{Automated Discovery and Optimization of 3D Topological Photonic Crystals}

\begin{abstract}
Topological photonic crystals have received considerable attention for their ability to manipulate and guide light in unique ways.
They are typically designed by hand based on careful analysis of their bands and mode profiles, but recent theoretical advances have revealed new and powerful insights into the connection between band symmetry, connectivity, and topology.
Here we propose a combined global and local optimization framework that integrates a flexible symmetry-constrained level-set parameterization with standard gradient-free optimization algorithms to optimize topological photonic crystals, a problem setting where the objective function may be highly non-convex and non-continuous.
Our framework can be applied to any symmetry-identifiable band topology, and we demonstrate its applicability to several prominent kinds of three-dimensional band topology, namely $\Gamma$-enforced nodal lines, Weyl points, and Chern insulators. 
Requiring no prior examples of topological photonic crystals or prior knowledge on the connection between structure and band topology,   
our approach indicates a path towards the automated discovery of novel topological photonic crystal designs.
\end{abstract}

\maketitle

%-----------------------
%----- MAIN MATTER -----
%-----------------------

% ------------------------------------
% ------------------------------------
\section{Introduction}

The past few decades have seen tremendous advances in optimization and inverse design techniques for nanophotonic components and, in particular, photonic crystals (PhCs)~\cite{jensen2010topology, molesky2018inverse, li2019topology}.
These advances have been spurred partly by the increase in computing capacity and methodological innovations, and partly by emerging micro- and nanofabrication capabilities that have significantly expanded the overall design space for device structures.
A central and persisting effort has focused on the optimization of PhCs with large photonic band gaps, with approaches including gradient-based~\cite{dobson1999maximizing,cox2000band,kao2005maximizing}, semi-definite programming (SDP)~\cite{men2010bandgap,men2014robust}, and gradient-free~\cite{goh2007genetic,yan2021photonic} techniques.
In more recent years, topological PhCs have emerged as a particularly exciting research direction, and with the more recent introduction of topological band theory to photonics~\cite{lu2014topological, ozawa2019topological, tang2022topological}, the PhC's topology has also emerged as a quantity of interest for design.
Here, we revisit the question of optimizing photonic band gaps in this new context, seeking the automated discovery and optimization of large, topologically nontrivial band gaps and well-isolated, robust topological degeneracies.

Topological PhCs are attractive due to their promised robustness for photonic devices and the plethora of associated exotic optical phenomena~\cite{lu2014topological, ozawa2019topological,tang2022topological}.
In contrast to their electronic counterparts, topological PhCs offer the exciting prospect of a flexible and tunable design space, with structural morphology subject only to the limits of fabrication constraints.
This design freedom has not only endowed photonics with the underlying technological promise of topological protection, but also positioned it as a promising platform for the fundamental exploration of spinless topological physics.
Exemplifying this, the first experimental realization of a Chern insulator without Landau levels---
the quantum anomalous Hall effect~\cite{haldane1988model, raghu2008analogs}---was achieved in a gyromagnetic 2D PhC~\cite{wang2009observation}.
Similarly, the first experimental observation of a Weyl point was achieved in a 3D high-index PhC~\cite{lu2015experimental}, simultaneously with its observation in TaAs~\cite{xu2015discovery, lv2015experimental}.
Gapless photonic topology has since grown to encompass a wide variety of phases, including unconventional Weyl points~\cite{jorg2022observation, shi2020spin, zhang2018double}, Dirac points~\cite{wang2016three, guo2019observation, wang2017type}, and nodal lines~\cite{lu2013weyl,yan2018experimental, gao2018experimental, xia2019observation, yang2020observation, park2021non, wang2021intrinsic, park2022nodal}.

These advances in photonic topology have been achieved largely by using analogy, physical intuition, and symmetry requirements.
Similarly, optimization of associated designs have depended mainly on parameter sweeps~\cite{xie2021optimization} and trial-and-error.
Although recent studies have applied topology optimization to topological PhCs~\cite{christiansen2019designing, zin2018topology, nussbaum2022optimizing}, these efforts have all relied on optimization of a continuous proxy objective to implicitly encourage topology-associated behavior (\eg power flow~\cite{christiansen2019designing}, directionality of Purcell enhancement~\cite{nussbaum2022optimizing}, or density of states~\cite{zin2018topology}) in lieu of explicitly enforcing a discrete topological constraint.
In addition to requiring a different formulation of this proxy for each topological effect, a significant downside of this approach is the frequent need for an initial candidate for optimization which is already topological~\cite{christiansen2019designing, nussbaum2022optimizing} to avoid designs that maximize the proxy objective but lack the intended band topology.
There are two significant challenges to explicitly enforcing band topology.
First, they are inherently discrete and discontinuous, rendering commonly-used optimization algorithms that assume differentiability nominally inapplicable.
Second, the conventional evaluation of topological invariants from band holonomy~\cite{alexandradinata2015thesis} (\ie parallel transport of wave functions, involving notions of band connections and curvatures) is computationally costly since they require simulating the full Brillouin zone (BZ) (\eg 2D Chern number) or a dense subset of it (\eg Berry phase or 3D Chern vector).

\begin{figure*}
    \centering
    \includegraphics[scale=.9]{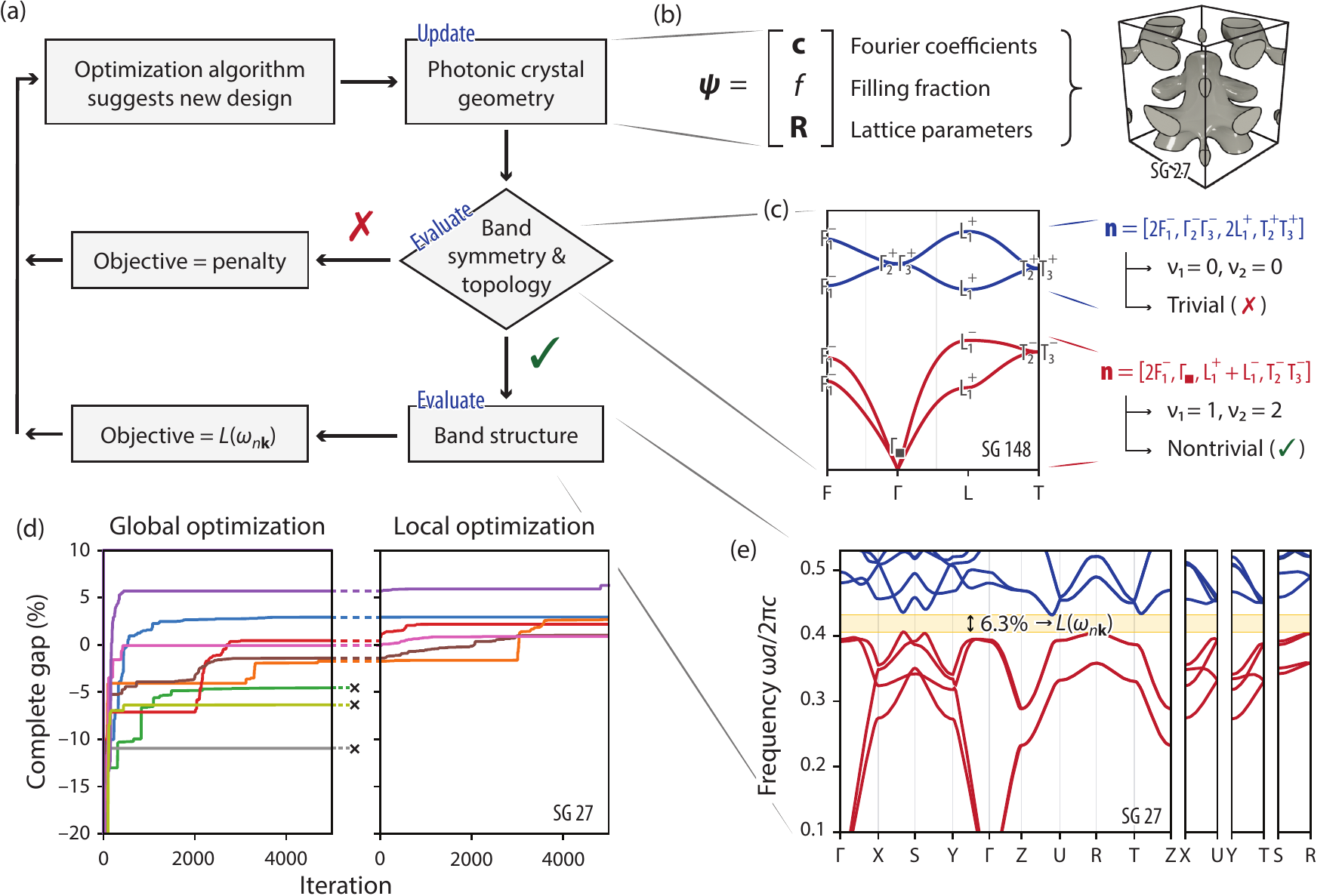}
    \caption{%
        \textbf{Overview of topological PhC optimization.}
        (a)~Flowchart of the optimization process in each iteration.
        Optimization continues until a convergence criterion or reaching a maximum number of iterations.
        (b)~The structure of the PhC is parameterized by a continuous vector consisting of the geometry coefficients (the Fourier sum coefficients), the filling fraction, and the unit cell lattice parameters.
        (c)~Symmetry-based tools can be used to calculate the band connectivity and topology from the high-symmetry (HS) $\kv$-points, providing a computationally efficient evaluation of the topology constraint (here, exemplified for a hypothetical PhC in SG 148).
        (e)~If the PhC is found to have the desired topological indicator as computed from the band symmetries, then the bandstructure is calculated along the HS $\kv$-lines (shown in purple in the inset).
        This example shows a complete bandgap highlighted in yellow.
        (d)~Optimization performance of complete bandgaps in SG 27 over multiple trials, as measured by the best value of the objective found so far as a function of iteration.
        Optimization consists of a 2-step process:
        global optimization with multiple trials, followed by
        local optimization using the best candidates from global optimization.
        Colors represent different trials from random initializations.}
    \label{fig:overview}
\end{figure*}

In this work, we propose to use a combination of gradient-free optimization algorithms, a symmetry-constrained level-set parameterization of the structure, and a computationally efficient symmetry-based evaluation of band topology to automate the discovery and optimization of novel topological PhCs.
An overview of the optimization process' elements is shown in \cref{fig:overview}, the flow of which is summarized in \cref{fig:overview}(a).
By using a level-set function expressed as a symmetry-constrained Fourier sum, we can parameterize even highly complex geometries using only a low-dimensional parameter space [\cref{fig:overview}(b)], which in turn makes both global and local gradient-free optimization algorithms feasible.
To efficiently evaluate the band topological constraints, we incorporate the recently introduced frameworks of topological quantum chemistry~\cite{bradlyn2017tqc} and symmetry indicators~\cite{po2017symmetryindicators, slager2017classification}, which have also been applied to PhCs~\cite{depaz2019engineering,christensen2022location}.
Furthermore, we use a global optimization algorithm in conjunction with a stochastic local optimization stage to escape local optima during optimization while also handling the non-continuous objective function stemming from the topology of the PhC.
We focus on 3D topological PhCs and present novel examples in three settings:
\begin{enumerate*}[label=(\roman*)]
\item $\Gamma$-enforced topological nodal lines,
\item ideal (\ie frequency-isolated) Weyl points in the interior of the BZ, and
\item photonic Chern insulators with chiral surface states.
\end{enumerate*}
In the last setting, we find a photonic Chern insulator with the largest known complete bandgap, and achieve this without explicitly relying on the supercell modulation technique that underlies earlier designs.
While we focus on a handful of concrete topological properties, our method can in principle be applied to any symmetry-identifiable band topology.

% ------------------------------------
% ------------------------------------
\section{Methods}

% ------------------------------------
\subsection{Photonic Crystal Parameterization}

A key choice in the optimization of any system is the parameterization, \ie the degrees of freedom. 
In the context of PhCs, this entails a parameterization of the dielectric function $\varepsilon(\rv)$ across the unit cell.
Historically, a widely employed approach has involved restricting the PhC structure to simple ``basic'' geometric shapes (\eg circles, squares, spheres, or cubes of dielectric material) requiring very few geometric parameters (\eg radius or width), which in turn enables optimization using gradient-free techniques or even plain parameter sweeps~\cite{joannopoulos2008molding}.
Such approaches have been surprisingly effective, achieving for example (trivial) complete three-dimensional bandgaps of nearly $35\%$ with an index contrast of 4 in a diamond lattice of spherical holes~\cite{ho1990existence}.
At the opposite end of the spectrum, topology optimization (referring to the optimization of the geometric and topological shape of a \emph{structure}; not to be confused with topology of \emph{bands}) is an extremely flexible approach that divides the design region into a grid of pixels or voxels, each representing a continuous design parameter.
The associated dimensionality of the design space can be very high, ranging from thousands to millions.
To make this optimization tractable, gradient-based algorithms~\cite{dobson1999maximizing,cox2000band} or subspace methods combined with SDP~\cite{men2010bandgap,men2014robust} have been used, enabling optimization of PhC bandgaps in both 2D and 3D.

To represent the permittivity $\varepsilon(\rv)$ over the coordinates $\rv$ of the unit cell, we adopt a level-set parameterization. Level-set functions are significantly more flexible than combinations of simple shapes while making the problem more tractable than voxel-based discretizations for gradient-free optimization algorithms.
Concretely, we introduce a level-set function $\phi(\rv)$ whose intersection with a level-set offset $\Delta$ determines the boundary between regions of permittivity $\varepsilon_1$ and $\varepsilon_2$:
\begin{equation}
    \varepsilon(\mathbf{r})
    =
    \begin{cases}
    \varepsilon_2 & \Re\phi(\mathbf{r}) > \Delta \\
    \varepsilon_1 & \Re\phi(\mathbf{r}) \leq \Delta
    \end{cases},
\end{equation}
Going forward, we will take $\varepsilon_1 = 1$, corresponding to vacuum or air, and denote $\varepsilon_2$ simply by $\varepsilon$.
The advantages of a level-set function is that it allows a relatively low-dimensional parameterization while retaining a versatile geometric design space.
The technique has been widely employed in photonic optimization, with level-set functions parameterized by spherical harmonics~\cite{miller2014fundamental}, Hamilton-Jacobi formulations~\cite{kao2005maximizing}, and eigenfunctions of a correlation function~\cite{yan2021photonic}.

Here, exploiting the periodic nature of PhCs, we parameterize the level-set function $\phi(\rv)$ by a finite Fourier summation of plane waves with spatial frequencies at the reciprocal lattice vectors $\{\Gv\} = \{n_1\Gv_1 + n_2\Gv_2 + n_3\Gv_3 \mid n_i\subset \mathbb{Z}\}$, \ie by:
\begin{equation}
    \label{eq:fourier-sum}
    \phi(\mathbf{r};\mathbf{c_\mathbf{G}}) = \sum_{\{\mathbf{G}\}}c_\mathbf{G} \e^{\iu\mathbf{G}\cdot\mathbf{r}},
\end{equation}
with complex expansion coefficients $\mathbf{c} = \{c_{\Gv}\}$.
The Fourier parameterization has a number of advantages:
\begin{enumerate*}[label=(\roman*)]
\item it automatically incorporates lattice periodicity,
\item imposing an upper cutoff $G_{\mathrm{max}}$ on the norm of included reciprocal lattice vectors translates to limiting the spatial variation of the permittivity profile (\ie small feature sizes) and approximates feature-size constraints, and
\item symmetry constraints can be straightforwardly incorporated which additionally translates to a reduction in the number of free expansion coefficients
\end{enumerate*}.

The third point---symmetry-constraints---is crucial in our context, given the close connection between symmetry and band topology that we aim to exploit.
We briefly summarize how symmetry constraints can be imposed on the Fourier coefficients, following Ref.~\citenum{christensen2022location}.
Concretely, the permittivity $\varepsilon(\rv)$ must be invariant under every symmetry operation $g$ in the space group (SG) $\mathcal{G}$, such that $g\varepsilon(\rv) = \varepsilon(g^{-1}\rv) = \varepsilon(\rv)$---or equivalently $\phi(g^{-1}\rv) = \phi(\rv)$ for every $g \in \mathcal{G}$.
This imposes the coefficient constraints $c_\Gv \e^{-\iu g\Gv\cdot\mathbf{w}} = c_{g\Gv}$, with $g$ expressed in Seitz notation $g=\{\mathbf{W}|\mathbf{w}\}$ with rotation part $\mathbf{W}$ and translation part $\mathbf{w}$.
These constraints impose a set of interrelations among the Fourier coefficients, linking $c_{\Gv}$ of symmetry-related reciprocal lattice vectors $\mathop{\mathrm{star}}(\Gv) = \{g\mathbf{G} \mid g\in\mathcal{G}\}$, \ie linking coefficients associated with distinct stars of $\Gv$.
Overall, this reduces the unconstrained Fourier summation in \cref{eq:fourier-sum} to a symmetry-constrained sum:
\begin{equation}
\label{eq:fourier-sum-symmetry}
    \phi(\mathbf{r};\mathbf{c_{\text{star}(\mathbf{G})}}) = \sum_{\{\text{star}(\mathbf{G})\}} c_{\text{star}(\mathbf{G})} \sum_{\mathbf{G} \in \text{star}(\mathbf{G})} c_\mathbf{G}^{\text{star}(\mathbf{G})} \e^{\iu\mathbf{G}\cdot\mathbf{r}}.
\end{equation}
Here, $c_\mathbf{G}^{\mathop{\mathrm{star}}(\mathbf{G})}$ denotes fixed, symmetry-determined coefficients among reciprocal lattice vectors from the same star while $c_{\mathop{\mathrm{star}}(\mathbf{G})}$ denotes the remaining free coefficients, each associated with a specific star.
We only need to optimize the free parameters $\mathbf{c_{\text{star}(\mathbf{G})}}$.
Note that the number of parameters thus depends on the SG: roughly, more symmetries (larger group order) translate to fewer free parameters.
Generally, these parameters are complex; in the presence of inversion symmetry, however, they can be restricted to be real parameters.

In practice, we limit the included stars of $\Gv$ to those that have a representative element with components $n_i \in \{0, \pm 1, \pm 2\}$.
In addition, we parameterize the level-set boundary using the filling fraction $f$ rather than the level-set offset $\Delta$, since the mapping between $\Delta$ and $f$ is monotonic and the filling fraction is more intuitive to interpret.
This choice is also motivated by the empirical observation that a filling fraction parameterization leads to better optimization performance (likely because uniform sampling of $\Delta$ and $c_{\mathop{\mathrm{star}}(\Gv)}$ produces structures with normally distributed filling fractions of small variance and mean close to 0.5, which is an undesirable bias).
We bound the values of $f$ and $c_{\mathop{\mathrm{star}}(\Gv)}$ to $f\in[0,1]$ and $c_{\mathop{\mathrm{star}}(\Gv)}\in[-1, 1]$. 
Altogether, the symmetry-constrained level-set Fourier parameterization of \cref{eq:fourier-sum-symmetry} allows a great deal of shape design freedom, as illustrated \eg in \cref{fig:overview}(b), while requiring only very few parameters (in the range 11--69 for the problems considered here).

We also include the unit cell's lattice vectors $\{\Rv_1, \Rv_2, \Rv_3\}$ as part of the overall structure specification, which can be parameterized (up to an immaterial overall rotation) by their lengths $\{a,b,c\}$ and mutual angles $\{\alpha = \angle(\Rv_2,\Rv_3), \beta = \angle(\Rv_3, \Rv_1), \gamma = \angle(\Rv_1,\Rv_2)\}$.
We exploit the scale-invariance of the dispersion-free Maxwell's equations to eliminate one of these parameters, setting $a = 1$ without loss of generality.
The remaining 5 parameters are typically constrained further by the choice of SG:
\eg SGs in the cubic crystal system (195--230) have no free lattice parameters whereas SGs in the monoclinic crystal system (3--15) have 3 free parameters ($b$, $c$, and $\beta$)~\cite{ITA}.
As a practical matter, we restrict the free lattice vector lengths and angles to $b, c \in [0.75, 1.25]$ and $\alpha, \beta, \gamma \in [90^{\circ}, 150^{\circ}]$ during optimization.
Intuitively, we expect that a nearly isotropic BZ will tend to host larger HS bandgaps; empirically, we observe that restricting $b, c$ to near unity indeed improves optimization convergence.

Unless otherwise specified, we adopt a scalar, fixed permittivity of $\varepsilon=16$, roughly representative \eg of silicon at visible frequencies or certain ceramic-filled plastics at microwave frequencies~\cite{lu2015experimental}.
The permittivity could in principle be included as an additional optimization parameter; in practice, however, bandgap optimization tends to automatically favor increased permittivity, rendering its optimization trivial.

% ------------------------------------
\subsection{Objective}

In the topological settings explored in this work, we wish to maximize the bandgap along high-symmetry (HS) $\kv$-lines in the BZ.
However, bandgap optimization is a notably difficult problem as the bandgap objective is not fully differentiable (especially at points of band crossing) and is a highly non-convex problem with numerous local optima~\cite{cox2000band,men2010bandgap}.
Earlier works were only able to optimize the bandgap if the structure used as an initial point for optimization already had at least an incomplete gap~\cite{dobson1999maximizing,cox2000band}.
To address this limitation, \citet{cox2000band} combine gradients with evolutionary algorithms to better search the entire parameter space, while \citet{men2010bandgap} transform the objective and constraints to improve the differentiability of the problem.
However, these methods are still prone to getting trapped in local maxima.
We note that using the SDP and subspace method developed by \citet{men2014robust} to maximize the bandgap in SG 13, we were not able to find a single example with a complete HS bandgap over dozens of trials.
Furthermore, in optimizing topological PhCs we introduce additional constraints in the form of the topological indicator of the bands of interest.
Topology is inherently a discrete quantity,and so the constraints are non-continuous, making this problem difficult to address using existing approaches.

To address these challenges, we propose an optimization framework shown in \cref{fig:overview}(a) which uses gradient-free optimization algorithms. 
This optimization framework allows us to (1) simultaneously optimize the lattice parameters as we are interested in SGs that do not have cubic lattices, (2) easily search the global space which is highly non-convex and has numerous local maxima, (3) handle non-differentiable objectives, and (4) incorporate non-continuous constraints corresponding to the topological indicators of the bands.

In particular, we choose the objective function $L(\omega_{n\kv})$---\ie the quantity we wish to maximize---as the relative bandgap:
\begin{equation}
\label{eq:objective-n}
    \Delta \omega_n = 2\frac{\min\limits_\mathbf{k} \omega_{n+1}(\mathbf{k}) - \max\limits_\mathbf{k} \omega_{n}(\mathbf{k})} {\min\limits_\mathbf{k} \omega_{n+1}(\mathbf{k}) + \max\limits_\mathbf{k} \omega_{n}(\mathbf{k})}
\end{equation}
where $\omega_n(\mathbf{k})$ is the $n$th band from the bottom. $L(\omega_{n\kv})$ can be in general a function of $\kv$ across the entire BZ.
However, to make optimization of 3D PhCs computationally efficient, we restrict the domain of $\kv$ to the HS $\kv$-lines as shown in \cref{fig:overview}(e), discretized into approximately 300 equally spaced $\kv$-points.

For some applications, such as $\Gamma$-enforced topology, $n$ is fixed.
However, for other applications where we are agnostic to the particular band and simply wish to maximize \emph{a} gap, we search over multiple bands by using the objective function:
\begin{equation}
\label{eq:objective-all}
    \Delta \omega = \max_n \Delta \omega_n
\end{equation}
where $n$ can vary from 1 up to an upper limit we set based on computational constraints.

In the context of topological PhCs, we want to constrain bands 1 through $n$ (where $n$ may be fixed or variable) to have a particular topological index. 
To enforce the topology of the bands, we check the topological index in each iteration as shown in \cref{fig:overview}(c), as symmetry indicators allow us to quickly calculate the index from only the HS $\kv$-points (the details of which are explained in the section on ``PhC Simulation'').
If the bands are indeed topological, we then calculate the entire band structure along the HS $\kv$-lines from which we set $L(\omega_{nk})= \Delta \omega_n$ or $L(\omega_{nk})= \Delta \omega$, as appropriate.
Otherwise, we assign a penalty term of $-2$ to the objective, as this is the minimum possible value of the relative bandgap.
Incorporating the constraint into the objective function allows us to use a variety of optimization algorithms that are not necessarily designed for nonlinear constraints.
The overall process in each iteration is summarized in \cref{fig:overview}(a)

% ------------------------------------
\subsection{Optimization}

As shown in \cref{fig:overview}(d), the optimization framework consists of 2 stages: a global optimization stage to search the entire parameter space, and a local optimization stage that takes the best candidates from the first stage and fine-tunes them within a small region in the parameter space.

The first stage of optimization uses a randomized variant of the DIRECT-L algorithm (which we will refer to as DIRECT-L-RAND) as implemented by NLopt~\cite{johnson2014nlopt,gablonsky2001locally}.
The randomized variant allows us to run the algorithm over multiple trials to take advantage of high-performance computing cluster resources. 
We also use a random initial point to further introduce diversity over different trials.
In principle, the non-randomized variants, DIRECT or DIRECT-L, can be used in place of DIRECT-L-RAND, although they are insensitive to the initial point and different trials will generally converge to the same solution.

\begin{figure*}
    \centering
    \includegraphics[scale=1]{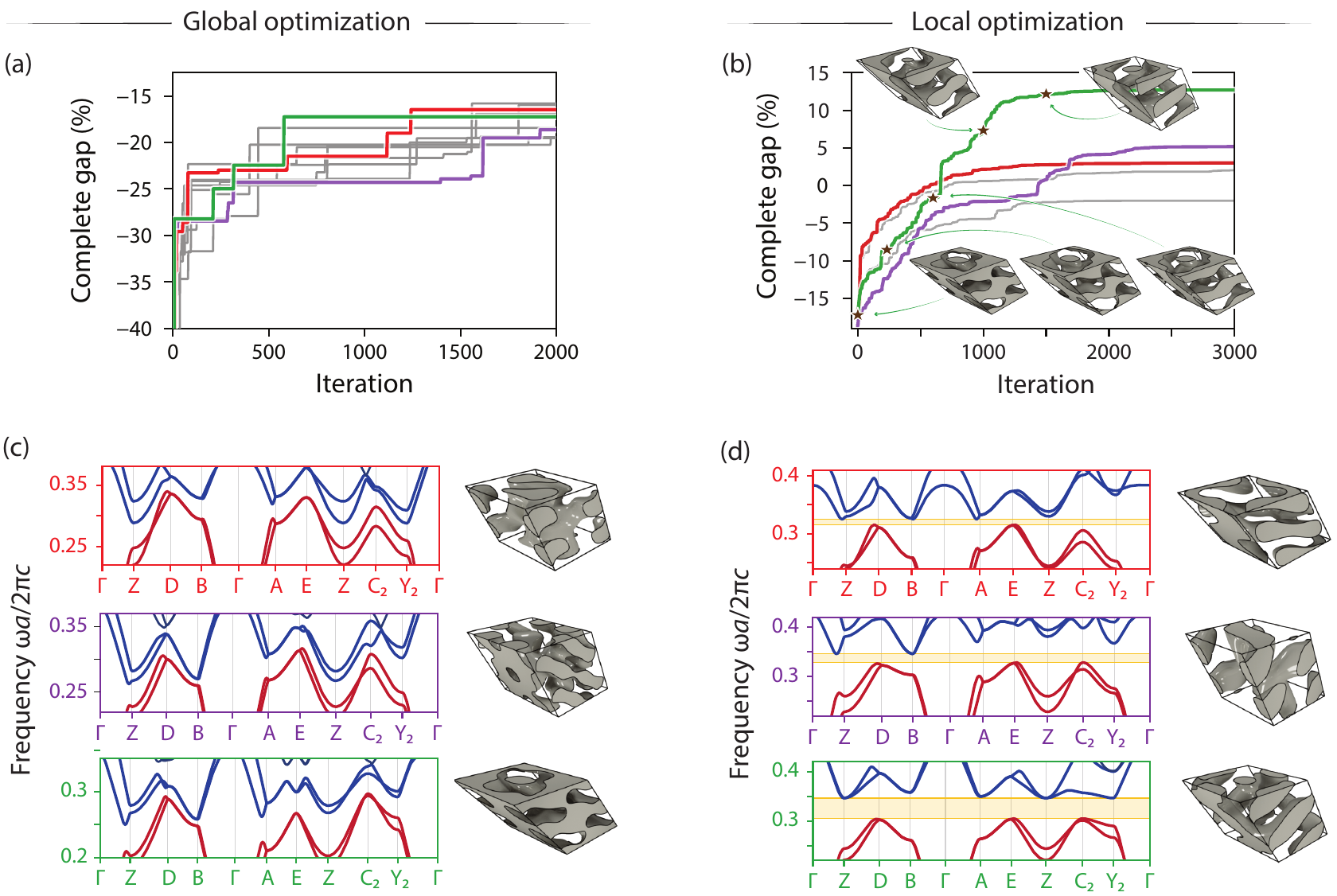}
    \caption{%
        \textbf{Global and local optimization of PhCs in SG 13 with $\Gamma$-enforced topology.}
        (a--b)~Evolution of the the complete HS gap, \ie the optimization objective, during the (a) global and (b) local optimization stages.
        Different curves correspond to different trials from random initializations.
        Of the 10 trials shown in (a), the 5 best trials are used as initialization for the local optimization stage.
        The three trials with best eventual local optimization result are color-highlighted (red, purple, and green).
        The unit cell evolution during local optimization for the green trial is shown at select iterations (stars).
        (c--d)~Band structures and unit cells for the color-highlighted (red, purple, and green) optimization curves at the final iteration of global (c) and local (d) optimization.
        }
    \label{fig:sg13-results}
\end{figure*}

As shown in \cref{fig:sg13-results}(a), the global optimization stage not only is able to quickly find candidates with the desired topological index, but also produces a variety of possible candidates from which we can start the local search.
The PhC geometries as well as their band structures are quite diverse, demonstrating the effectiveness of stochastic global exploration.

\begin{figure}
    \centering
    \includegraphics[scale=0.9]{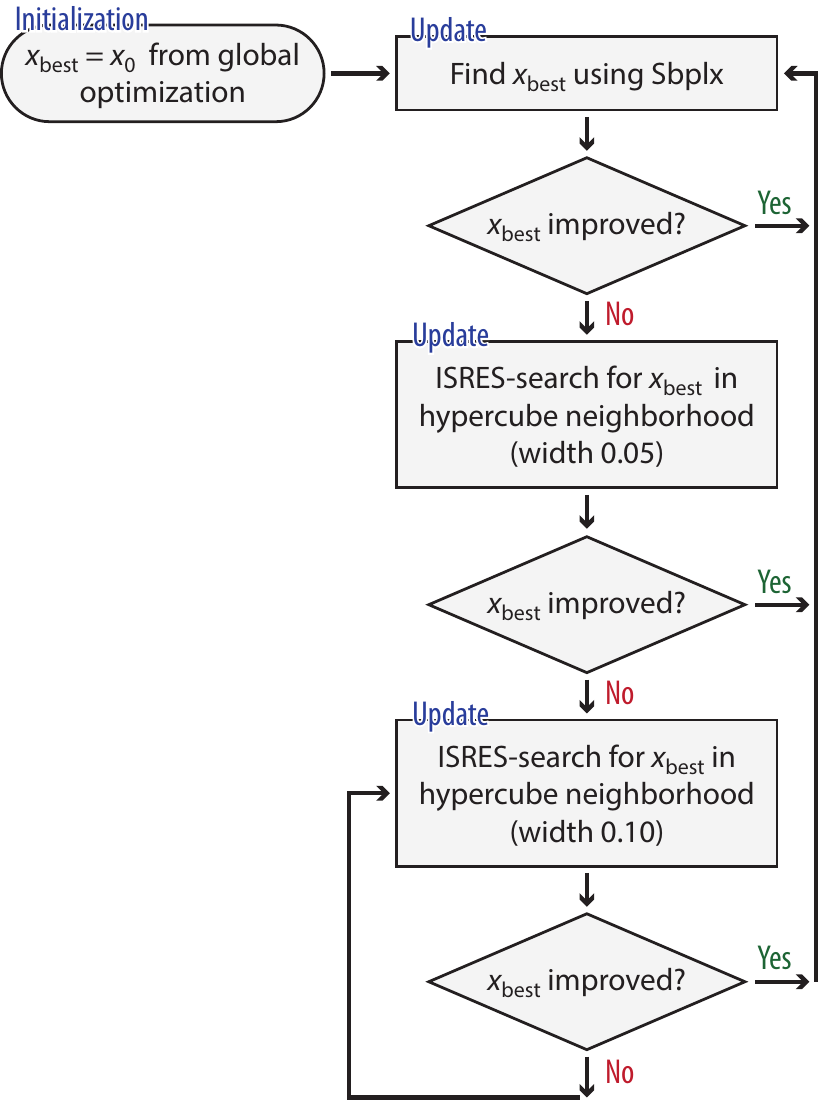}
    \caption{%
        \textbf{Flowchart of the local optimization algorithm.}
        Due to the non-differentiable objective function, Sbplx is subject to premature convergence.
        To ameliorate this, we periodically switch to ISRES to escape local maxima and continue optimization.}
    \label{fig:local-opt}
\end{figure}

The second stage is a local optimization to refine the best candidates found from global optimization.
We primarily use the Sbplx algorithm, a variant of Nelder-Mead, as implemented by the NLopt library~\cite{johnson2014nlopt,rowan1990functional}.
The challenge with applying popular local optimization algorithms to the problem of topological PhCs is that they often model the objective and constraints inside trust regions using linear approximations and thus require differentiability of both the objective and constraints.
However, the discrete nature of topology means that the constraint (or objective) is non-differentiable, which leads to the failure modes in the optimization algorithm, including getting stuck in non-feasible regions of the parameter space, decreasing the step size to the point where it no longer improves the objective, or even diverging and jumping to a point far away from the initial point.
Thus, to avoid the local optimization prematurely converging, we periodically restart the search algorithm starting from the best point found so far to reset the trust region and step size.
We also occasionally switch to ISRES~\cite{runarsson2005search, johnson2014nlopt}, a stochastic evolutionary-based global optimization algorithm, to escape local maxima.
The bounds of ISRES are set to be a small hypercube around the best point found so far so that it behaves like a stochastic local optimization.
This process is shown in \cref{fig:local-opt}.

As seen in \cref{fig:sg13-results}(b), the local optimization stage is effective in quickly and significantly increasing the bandgap.
The vast difference in performance between global and local optimization stages indicates the difficulty of the bandgap problem and suggests that the set of feasible parameters that support complete HS bandgaps is small relative to the parameter space.
The diversity of the results from global optimization and the fact that the different trials do not converge to the same structure during local optimization also demonstrates that the bandgap problem likely has many local optima, many of which do not have complete HS bandgaps.
Thus, multiple trials with different random initializations are necessary to find the globally optimal structure.

Our framework is similar to multi-start local optimization algorithms which apply local optimization algorithms multiple times from different points to find the global maxima in non-convex problems.
The difference is that we apply multiple trials of stochastic global optimization to pick the starting points for local optimization rather than randomly sampling the starting points.
This is necessary due to (1) the extremely large number of local maxima in the parameter space and (2) the topology constraints.
Only a small fraction of randomly sampled points obey the desired topology constraints (typically $<10\%$ and often $<1\%$), so the global search is necessary to find valid starting points with the desired topology.
Thus, unlike prior work that optimizes topological PhCs from a known topological structure~\cite{xie2021optimization, nussbaum2022optimizing}, our method is able to automatically discover new topological PhCs.

\subsection{PhC Simulation}

We use the MIT Photonics Bands (MPB) software to solve for the eigenmodes of the PhC unit cell and calculate the band structure and associated symmetry eigenvalues~\cite{johnson2001block}.
Specifically, we use a $16\times 16 \times 16$ spatial resolution during optimization and a $32\times 32 \times 32$ resolution when calculating the full BZ.
For the symmetry-based evaluation of band topology, we use the Julia packages Crystalline.jl~\cite{christensen2022location}, which implements the methodology of topological quantum chemistry~\cite{bradlyn2017tqc} and symmetry indicators~\cite{slager2017classification, po2020symmetry}; and MPBUtils.jl, which implements additional MPB-specific functionality~\cite{MPBUtils.jl}. The choice of HS $\kv$-paths is not unique for a particular SG; we use the SeeK paths as defined by \citet{hinuma2017band} and as implemented by Brillouin.jl~\cite{Brillouin.jl}.

We briefly summarize the salient ideas of the symmetry-based evaluation of band topology (see, \eg Refs.~\citenum{cano2021band} and \citenum{po2017symmetryindicators} for existing reviews of the general methodology and Ref.~\citenum{christensen2022location} for its adaptation to PhCs).
Given a selection of bands $\{n\}$, we first compute the band symmetry eigenvalues $\langle \Ev_{n\kv}|g\Dv_{n\kv}\rangle$ at the special $\kv$-points of the BZ, using the Bloch eigenmodes of the $\Ev$- and $\Dv$-field, over the symmetry operations $g$ of each $\kv$-point's little group (\ie the operations that leaves the $\kv$-point invariant, modulo reciprocal lattice vectors).
From the set of symmetry eigenvalues, we next compute the combination of irreducible representations (irreps) that the bands transform as at each $\kv$-point.
The overall information is aggregated into an integer-valued symmetry vector:
\begin{equation}
    \nv
    =
    [n_{\kv_1}^{1}, n_{\kv_1}^{2}, \ldots, n_{\kv_N}^{1}, n_{\kv_N}^{2}, \ldots, \mu],
\end{equation}
where $n_{\kv_i}^{\alpha}$ denotes the number of bands in the selection that transform as the $\alpha$th irrep $D_{\kv_i}^{\alpha}$ of the little group at $\kv_i$, and $\mu$ denotes the total number of bands in the selection. 

For a given PhC, we build up a set of symmetry vectors iteratively, starting from the lowest bands.
For each candidate symmetry vector, we test its consistency against the set of compatibility relations imposed by the SG---\ie constraints on how band symmetries can connect across the Brillouin zone~\cite{bouckaert1936theory,kaxirax2019quantum}---and if incompatible, additional bands are included until a compatible band grouping is found.
Each such compatible symmetry vector signals a set of bands which are (pointwise) gapped from all other bands along all HS $\kv$-lines.
The associated assignment of topology to each such symmetry vector can be achieved using the formalism of symmetry indicators~\cite{po2017symmetryindicators, slager2017classification} and topological quantum chemistry~\cite{bradlyn2017tqc}.
In brief, for each SG, a choice of a set of trivial generators $\{\mathbf{a}_i\}$ and a set of nontrivial generators $\{\mathbf{b}_j\}_{j=1}^{d^{\text{BS}}}$ exists such that any compatible vector can be expressed as their integer combination according to~\cite{song2018diagnosis}:
\begin{equation}\label{eq:symmetryindicator}
    \nv = \sum_i c_i \mathbf{a}_i + \sum_j \nu_j\mathbf{b}_j,
\end{equation}
where $c_i\in \mathbb{Z}$ are integer coefficients and $\nu_j \in \mathbb{Z}_{\lambda_j}$ are coefficients from the ring of integers modulo $\lambda_j$, \ie from $\mathbb{Z}_{\lambda_j} = \{0, 1, \ldots, \lambda_j-1\}$.
Here, $\lambda_j$ are the minimal integers that allow an integer-coefficient expansion of $\lambda_j\mathbf{b}_j$ in $\{\mathbf{a}_i\}$ (\ie $\lambda_j\mathbf{b}_j$ is trivial from the perspective of symmetry).
The symmetry indicator, which characterizes the symmetry-identifiable topology, is precisely the expansion coefficients $\nu_i$:
that is, we associate with every $\nv$ the symmetry indicator $\boldsymbol{\nu} = (\nu_1, \ldots, \nu_{d^{\text{BS}}})$ which is an element of the indicator group $\mathbb{Z}_{\lambda_1}\times\ldots\times\mathbb{Z}_{d^{\text{BS}}}$.
Nontrivial band topology corresponds to nonzero $\boldsymbol{\nu}$ and symmetry-identifiable topology consequently exists only in SGs with indicator group different from $\mathbb{Z}_1$.
In practice, we compute the compatibility relation tests as well as the symmetry indicators using a procedure described elsewhere~\cite{elcoro2020application, christensen2022location}.
A sketch of the application of the framework to a hypothetical PhC in SG 148 (R$\overline{\text{3}}$) with indicator group $\mathbb{Z}_2\times\mathbb{Z}_4$ is shown in \cref{fig:overview}(c).

We emphasize that the computational efficiency of this symmetry-based approach is crucial to our optimization technique.
With it, we can evaluate both the band connectivity and topology using just a handful of special $\kv$-points (6--8 for the SGs considered here); without it, this would require simulating a densely discretized subset of the BZ, rendering the evaluation of the penalty term computationally prohibitive.

% ------------------------------------
% ------------------------------------
% ------------------------------------
\section{Results and discussion}

This section presents a number of optimized PhCs with nontrivial topology, including $\Gamma$-enforced nodal lines, Weyl points, and Chern insulating gaps.

% ------------------------------------
\subsection{\texorpdfstring{$\Gamma$}{Gamma}-enforced topological nodal lines}
\label{sec:nodal-results}

We start by considering the optimization of PhCs exhibiting so-called $\Gamma$-enforced topological nodal lines,
a recently introduced type of photonic band topology that can be inferred solely from the number of bands below the first gap along the HS $\kv$-lines of the BZ~\cite{christensen2022location}.
In the presence of time-reversal symmetry, this band topology can exist in six SGs (13, 48, 49, 50, 68, and 86) and arises as a direct result of a uniquely photonic zero-frequency polarization singularity at the $\Gamma$-point.
It is associated with an apparent bandgap between bands 2 and 3 along HS lines of the BZ which, however, vanishes along nodal lines in the interior of the BZ.
Conceptually, it is analogous to filling-enforced topology~\cite{po2016filling}, \ie topology that can be inferred from band connectivity alone.
Here, we focus on optimizing the gap along the HS $\kv$-lines (which we denote as the ``HS bandgap'') with the aim of obtaining ``ideal'' frequency-isolated nodal lines. 
We explore each of the six SGs that support $\Gamma$-enforced nodal lines.
In all cases, $\Gamma$-enforced topology occurs only between bands 2 and 3, so we use the objective in \cref{eq:objective-n} with $n=2$.

Running our optimization algorithm for each of these six SGs, across 10--20 global trials and 5000 subsequent local iterations for the top 3--5 global candidates, we find $\Gamma$-enforced topological designs in every SG.
However, only in SG 13 (P2/c; generated by inversion and a glide operation $\{2_{010}|0,0,\tfrac{1}{2}\}$ in the monoclinic crystal system) are we able to find full band gaps across the HS $\kv$-path.
Notably, SG 13 is a (maximal) subgroup of the remaining SGs: the inability to obtain HS bandgaps in these supergroups suggests that their additional symmetry constraints act counter to gap formation.
Additionally, as shown in \cref{fig:sg13-results}, the optimization framework is able to find full HS bandgaps in nearly half of the trials.

\begin{figure}
    \centering
    \includegraphics[width=\columnwidth]{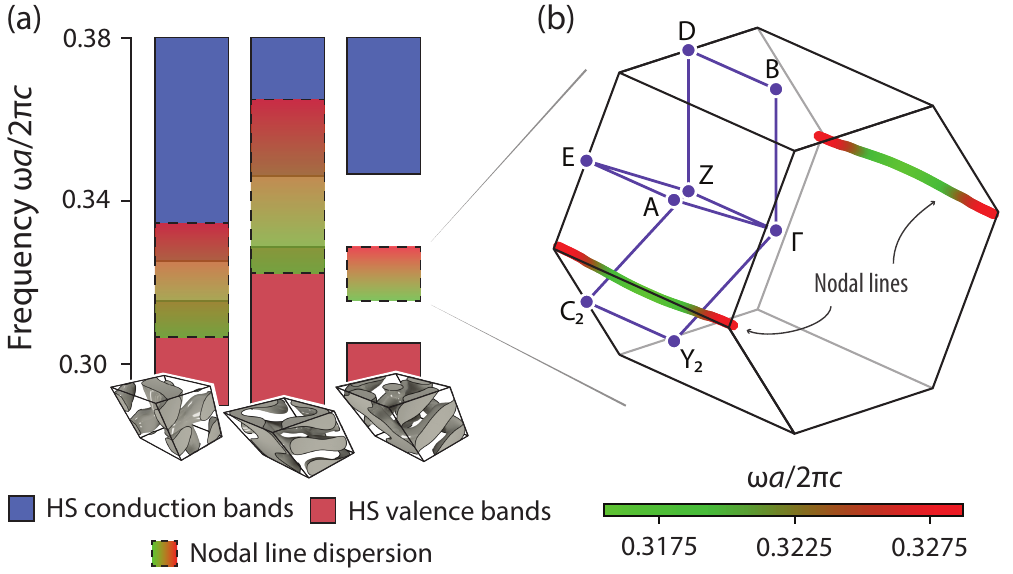}
    \caption{%
        \textbf{Nodal line location and dispersion in optimized candidates of SG 13.}
        (a)~Nodal line frequency dispersion compared to the HS bands. The right-most candidate exhibits a maximally-isolated nodal line.
        (b)~The nodal lines (color-coded by frequency) in the BZ (outlined in black) run along the $k_y$ direction.
        Purple lines represent the HS $\kv$-lines over which the bandgap is maximized
      }
    \label{fig:sg13-nodal}
\end{figure}

The best candidate we have found in SG 13 is presented in the bottom-most panel of \cref{fig:sg13-results}(b), which contains a $12.7\%$ HS gap between bands 2 and 3.
We can see that the structure converges to what resembles a double gyroid structure, a well-known design for complete bandgaps~\cite{lu2013weyl}.
The symmetry indicator of the bottom 2 bands is $(\nu_1,\nu_2) = (1, 1)$, indicating the presence of a pair of nodal lines, each protected by a $\pi$-Berry phase~\cite{song2018diagnosis}.
Performing a full BZ computation over a dense $\kv$-grid, we see a pair of nodal lines running along the $k_y$ axis in \cref{fig:sg13-nodal}(b), consistent with the predictions by \citet{song2018diagnosis}.
Unsurprisingly, the nodal lines are far away in the BZ from the HS $\kv$-lines along which we maximized the bandgap, since the band crossings that form nodal lines are antithetical to the optimization objective.
In other words, optimizing the HS bandgap pushes the nodal line away from the HS $\kv$-lines.

The nodal line is frequency dispersive, exhibiting a relative $4.2\%$ variation across its span.
While there exist isofrequency contours around the nodal line at the nodal line frequencies due to its dispersion, there do not exist any nodal line frequency pockets elsewhere in the BZ, making the nodal line well isolated in frequency.
Ideally, we would like to also minimize the frequency dispersion of the nodal line~\cite{yan2018experimental}, as it is necessary to realize the characteristic linear DOS dispersion of nodal lines.
However, this would require formulating an objective that is a function of the entire BZ rather than just the HS $\kv$-lines, which is computationally prohibitive; we leave such efforts to future work.

Because $\Gamma$-enforced topology is diagnosable from the connectivity alone (\ie if the appropriate bands are connected, they must be topological), optimization methods such as the SDP solver~\cite{men2014robust} that were not formulated specifically for topological PhCs can be applied for this setting without requiring additional constraints. 
However, starting from random initializations, the SDP solver was not able to find any complete HS gaps in SGs 13, 48, and 68 over multiple trials and over a range of lattice bases.
The difficulty in finding bandgaps points towards the nonconvexity of the problem and the necessity of global optimization methods.

\begin{figure}
    \centering
    \includegraphics[width=\columnwidth]{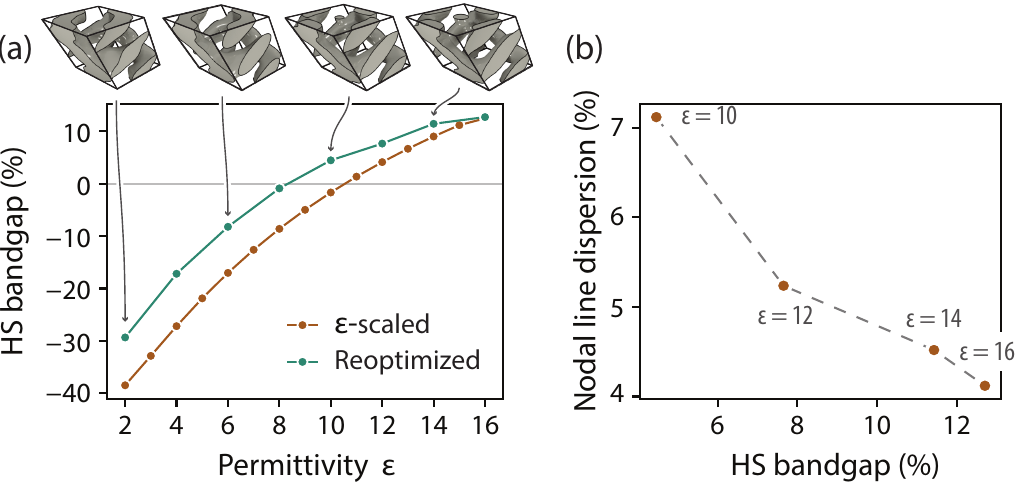}
    \caption{%
      \textbf{Permittivity dependence of nodal line properties.}
      (a)~We compare the HS gap attainable by simply scaling the permittivity of an original design (brown lines) versus the HS gap attainable by re-optimizing the structure at the new target permittivity (green lines), using the original structure as initialization point.
      The starting design is that associated with \cref{fig:sg13-nodal}(b), originally optimized at $\varepsilon=16$.
      Substantial improvements are attainable by re-optimizing.
      (b)~Nodal line frequency dispersion as a function of the HS bandgap for the re-optimized structures, showing a monotonic decrease of nodal line dispersion with HS gap size.}
    \label{fig:sg13-varyeps}
\end{figure}

\paragraph{Optimization under variable permittivity}
To study the minimum index required to support an ideal nodal line, we take the best candidate (which has a $12.7\%$ HS gap) and calculate the band structure as a function of index for the optimized structure in \cref{fig:sg13-varyeps}.
As expected, the HS bandgap increases monotonically with the index contrast.
The same structure has a HS bandgap of 0 at a relative permittivity of $\varepsilon \approx 10.6$.
Alternatively, if we take the same structure and re-optimize it using our local optimization algorithm in \cref{fig:local-opt}, we can further improve the bandgap at a fixed permittivity, decreasing the minimum permittivity required for a complete HS bandgap to $\varepsilon \approx 8.3$.
Looking at the nodal line for the optimized structures at different permittivities in \cref{fig:sg13-varyeps}(b), the nodal line frequency dispersion scales inversely with the attainable HS bandgap.
In other words, the nodal line spans a greater fraction of the bandgap in frequency as permittivity decreases, remaining isolated inside the bandgap down to $\varepsilon=12$.

Many previous proposals for PhCs containing nodal lines either require metallic materials or very high index-contrast dielectrics~\cite{lu2013weyl,park2021non,park2022nodal}.
Our work paves the way for demonstrating nodal lines in PhCs with lower index, potentially enabling experimental realization in a wider range of frequencies.

% ------------------------------------
\subsection{Weyl points}

\begin{figure*}
    \centering
    \includegraphics[width=\textwidth]{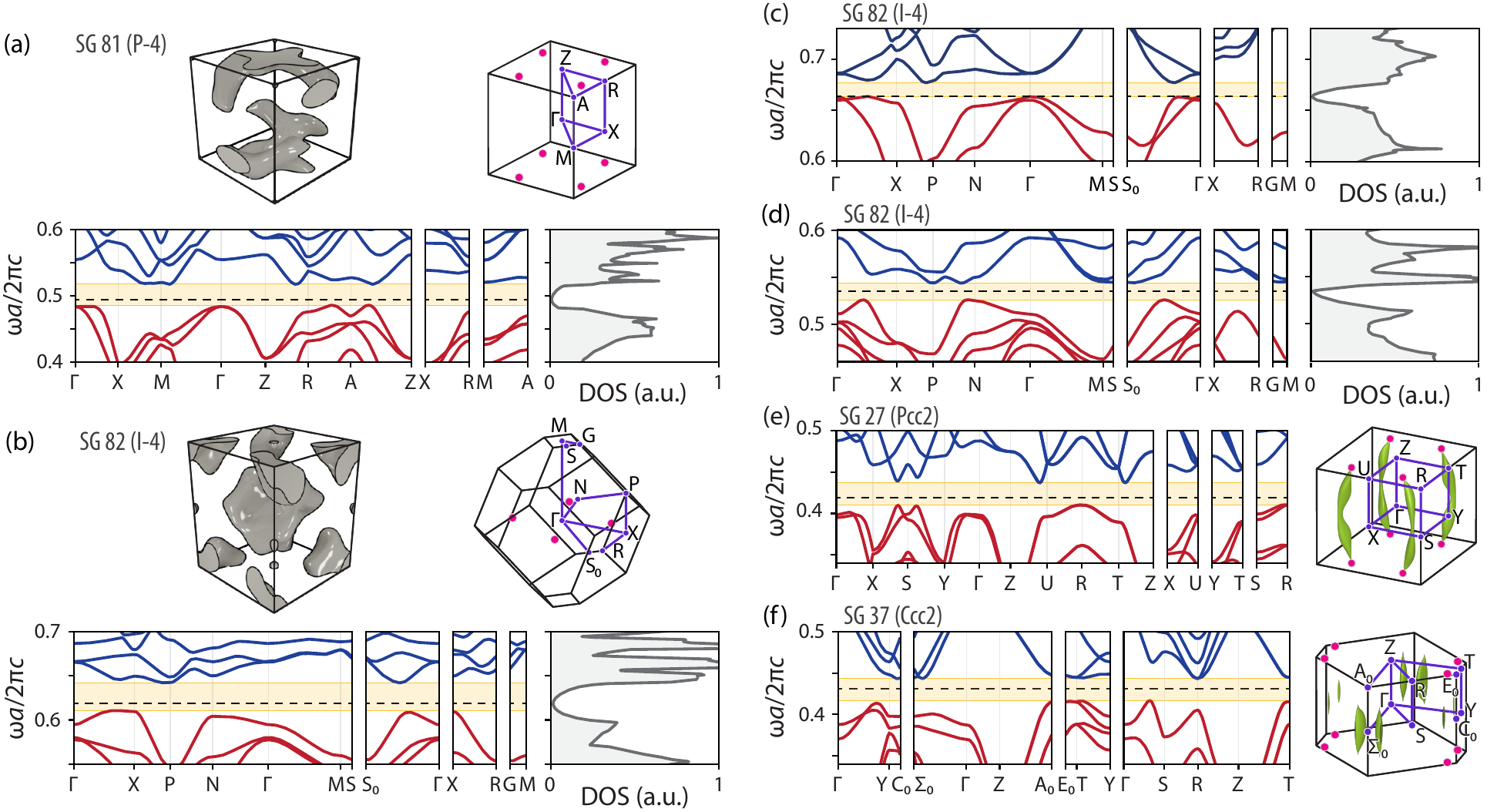}
    \caption{%
        \textbf{Weyl point optimization in noncentrosymmetric SGs.}
        Selected results for optimized PhCs designs with Weyl points in SGs 27, 37, 81, and 82, visualized by their dispersions along HS $\kv$-lines (red lines, nontrivial valence bands; blue lines, conduction bands; yellow shading, HS gap; dashed black line, Weyl point frequency), density of states (DOS), and Weyl point locations in the BZ.
        (a--b)~Optimized PhC Weyl point designs in SGs 81 and 82: the unit cells of each design are shown (for SG 82, we show the conventional unit cell, not the primitive).
        The BZs feature 4 ideal Weyl points in the $k_z=\pi/c$ plane (with equivalent copies in the $k_z=-\pi/c$ plane) in SG 81 and in the $k_z=0$ plane in SG 82, without intersecting Fermi pockets.
        The density of states (DOS) exhibits a parabolic frequency dependence at the Weyl point frequency.
        (c,d)~Optimized results in SG 82 with HS bandgaps, but non-ideal Weyl points. 
        (c)~The Weyl point frequency lies on the valence HS band edge, preventing frequency-isolation.
        (d)~Although the Weyl point frequency lies in the center of the HS bandgap, the DOS is asymmetric and non-parabolic, due to conduction band minima right above the Weyl point frequency that do not intersect the considered HS $\kv$-lines.
        (e-f)~Optimized PhC designs in SGs 27 and 37.
        In both cases, the Weyl points overlap spectrally with large Fermi pockets that extend over the interior of the BZ without intersecting any HS $\kv$-line.
        }
    \label{fig:weyl}
\end{figure*}

Next, we pursue designs of PhCs with frequency-isolated Weyl points, also known as ideal Weyl points.
Weyl points can arise in three different regions of the BZ:
\begin{enumerate*}[label=(\roman*)]
    \item at HS $\kv$-points, stabilized by spatial or time-reversal symmetries~\cite{manes2012existence, yu2022encyclopedia},
    \item along HS $\kv$-lines, corresponding to the intersection of bands transforming as different irreps~\cite{watanabe2018space, fang2020diagnosis}, and
    \item at generic $\kv$-points in the interior of BZ, stabilized by a nontrivial symmetry indicator~\cite{song2018diagnosis}.
\end{enumerate*}
To date, proposed designs of frequency-isolated ideal photonic Weyl points fall in the first two classes.
Here, we seek designs in the latter class, \ie Weyl points at generic momenta.
As before, we use the bandgap metric along HS $\kv$-lines as a proxy for isolating the Weyl points from other bulk bands.
Additionally, because we are agnostic to the number of bands below the gap, we allow the objective $L(\omega_{kn})$ to look over multiple bands, as in \cref{eq:objective-all}.

% --------------
\paragraph{Space group constraints}
\citet{song2018diagnosis} showed that for spinless, time-reversal invariant particles in non-centrosymmetric crystals (\ie lacking inversion centers), all symmetry-indicated nontrivial band topology is associated with Weyl points in the interior of the BZ.
We therefore initially restrict our attention to the 12 non-centrosymmetric SGs with nontrivial indicator group.
However, only a subset of these groups are suitable for finding well-isolated Weyl points since the Nielsen--Nimomiya theorem requires Weyl points to occur in pairs of opposite chirality, which in general do not need to have the same frequency~\cite{vanderbilt2018berry}.
However, the existence of additional symmetries, \eg mirrors or rotations, can map opposite-chirality Weyl points onto each other, thereby pinning them to the same frequency and allowing ideal single-frequency Weyl points.
Consulting the results of Ref.~\citenum{song2018diagnosis} for each SG individually, we find that only six SGs (27, 37, 81, 82, 103, and 184) can host ideal, symmetry-indicated Weyl points.
We perform global and local optimizations for PhCs in each of these SGs, and find designs with HS bandgaps in SGs 27 (Pcc2), 37 (Ccc2), 81 (P$\overline{\text{4}}$), and 82 (I$\overline{\text{4}}$).
SGs 27 and 37 are centering-variants, both generated by a 2-fold rotation and an orthogonal glide plane; similarly, 81 and 82 are centering-variants, both generated by 4-fold rotoinversion.

\paragraph{Ideal Weyl point designs}
We show results for a range of optimized designs from these SGs in \cref{fig:weyl}.
In SGs 81 and 82 we find several designs with ideal Weyl points and without any intersecting trivial bulk bands, \ie without any ``Fermi pockets''.
\Cref{fig:weyl}(a) highlights one such design from SG 81.
The band structure has a $6.3\%$ HS gap between bands 4 and 5, with Weyl points closing the gap in the interior of the BZ.
The ideality of the Weyl points is also clearly exhibited in the density of states (DOS), calculated using the tetrahedron method implementation from Ref.~\citenum{10.1088/2040-8986/aaae52}, which displays the characteristic parabolic dependence $\mathop{\mathrm{DOS}}(\omega)\propto(\omega-\omega_0)^2$ associated with ideal Weyl points at frequency $\omega_0$.
To determine the Weyl point locations, we perform a full BZ dispersion calculation and identify the Weyl points with the $\kv$-points where the conduction--valence frequency difference vanishes.
We observe 4 Weyl points in the BZ at $k_z=\pi/c$ (8 are shown; note, however, that the $k_z=\pm\pi/c$ planes are equivalent), consistent with expectations for a symmetry indicator $(\nu_1, \nu_2) = (0, 1)$~\cite{song2018diagnosis}.
Similar to the nodal line optimization, the Weyl points are pushed away from the HS $\kv$-lines in the $k_z=\pi$ plane, since the Weyl points represent a band-closing point which, if near a HS $\kv$-line, would act counter to the HS gap metric.

We have also found an example of a well-isolated, ideal Weyl point in SG 82, as shown in \cref{fig:weyl}(b).
The bandstructure displays a gap of $5.0\%$ between bands 6 and 7, and the DOS drops to 0 at the Weyl point frequency as expected.
In this SG, the Weyl points are fixed to the $k_z=0$ plane in the BZ.
Again, the Weyl points are pushed away from the HS $\kv$-lines by the optimization process.

\paragraph{Limitations of the high-symmetry bandgap metric}
While our optimization procedure has discovered several designs for ideal Weyl points, the HS bandgap metric is not infallible.
Concretely, even when large complete HS bandgaps are achieved, the corresponding Weyl points may not necessarily be truly frequency-isolated.
We identify two related failure modes.
First, the success of the HS bandgap metric hinges on an assumption of minimally varying dispersions between HS lines, \ie on the absence of non-Weyl-related local optima of the dispersion at generic $\kv$-points.
Although this is the tacit assumption underlying all the visualizations of the band structure along HS lines, it is well-known that this assumption can be violated~\cite{harrison2007occurence, craster2012dangers, maurin2018probability}.
Second, and relatedly, the frequency-location of the Weyl points relative to the HS bandgap is not pinned; specifically, although the \emph{existence} of Weyl points is guaranteed by topology, their frequency \emph{location} is not guaranteed to fall near the center of the HS bandgap.

\Cref{fig:weyl}(c-g) summarizes a few examples of these failure modes, where despite obtaining a large HS gap, the optimization procedure does not yield truly ideal Weyl points.
\Cref{fig:weyl}(c,d) shows the dispersions and DOS for two designs in SG 82 with bandgaps of $2.0\%$ and $3.3\%$, respectively.
In \cref{fig:weyl}(c) the Weyl point lies on the edge of the HS gap, illustrating the second failure mode mentioned above.
Despite this, the DOS still exhibits an approximately parabolic dependence near the Weyl point and vanishes exactly at the Weyl point frequency.
Even when the Weyl point occurs near the center of the HS gap, it may still be non-ideal, as illustrated by the design candidate considered in \cref{fig:weyl}(d):
the DOS is both asymmetrical and distorted from the expected parabolic shape due to the conduction band dropping close to, but not intersecting with, the Weyl point frequency elsewhere in the BZ, in illustration of the first failure mode mentioned earlier.

A more pronounced type of this failure mode, \ie spectral intersection with Fermi pockets despite an otherwise ``centered'' Weyl point, was widespread in the optimized designs found in SG 27 and 37.
Exemplifying this, \cref{fig:weyl}(e) shows an optimized PhC in SG 27 with a HS gap of $6.3\%$, and \cref{fig:weyl}(f) shows a PhC in SG 37 with a HS gap of $6.2\%$.
In both settings, the Weyl points are pinned to the $k_z=\pi$ plane (and, equivalently, $k_z=-\pi$ plane), and the Weyl point frequency lies in the middle of the HS gap.
Despite this, neither case exhibits ideal Weyl points; instead, the Weyl points are occluded by Fermi pockets that reside in the interior of the BZ, \ie there exist local optima of the bands inside the BZ.
We visualize these Fermi pockets in \cref{fig:weyl}(e,f) by plotting the isofrequency contours at the Weyl point frequency $\omega_0$.

Overall, the HS bandgap metric serves as a meaningful and largely effective proxy for optimizing Weyl points, as evidenced by our findings of ideal Weyl point designs in SGs 81 and 82.
Overcoming the limitations of the HS gap metric is in principle possible, \eg by replacing it by a full BZ gap metric (at considerable computational cost) or by supplementing the HS paths with additional paths, either from the outset or adaptively.
The DOS, which can be computed efficiently using the tetrahedron or Gilat--Raubenheimer methods~\cite{10.1088/2040-8986/aaae52}, could potentially also serve as an effective proxy.

% ------------------------------------
\subsection{Chern insulators}

\begin{figure*}
    \centering
    \includegraphics[width=\textwidth]{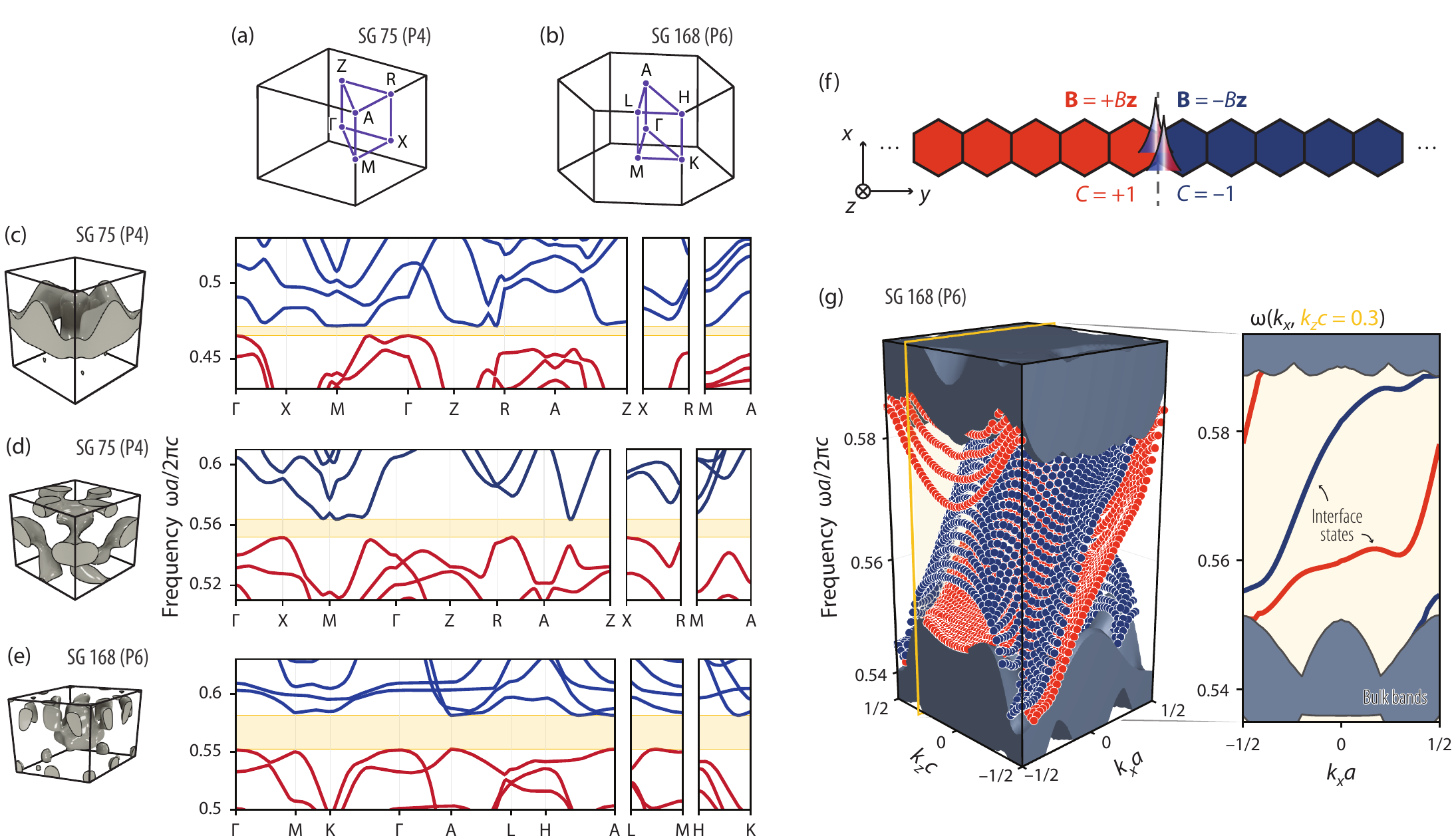}
    \caption{%
        \textbf{Optimization of photonic Chern insulators.}
        BZs and HS $\kv$-lines for 
        (a)~SG 75 and 
        (b)~SG 168, respectively.
        PhC structure and band structure (complete gap highlighted in yellow) for (c)~SG 75, $C = 1$, (d)~SG 75, $C = 2$, and (e)~SG 168, $C = 1$.
        (f)~Supercell calculation for the SG 168 candidate in (e) to calculate surface states.
        The supercell consists of $1\times16\times1$ unit cells.
        At the interface, the sign of the applied applied magnetic field $\mathbf{B}$ is flipped: \ie the orange units cells have $\mathbf{B} = +B\hat{\mathbf{z}}$ ($C = +1$) and the blue unit cells have $\mathbf{B} = -B\hat{\mathbf{z}}$ ($C = -1$).
        (g)~Surface state dispersion at the interface, extracted from supercell calculation (projected bulk states indicated in gray).}
    \label{fig:chern}
\end{figure*}

As a final showcase, we pursue designs of time-reversal broken 3D PhCs with nonzero Chern numbers, \ie photonic 3D Chern insulators.
There have been numerous proposals~\cite{raghu2008analogs, wang2008reflection, ao2009oneway, skirlo2014multimode, liu2012oneway} and associated experiments~\cite{wang2009observation, skirlo2015experimental, poo2011experimental, yang2013experimental, yin2013experimental} of photonic 2D Chern insulators.
The 3D generalization, however, has seen substantially less development, with only two designs known to us~\cite{lu2018topological, devescovi2021cubic}.
Both of these designs require supercell modulations, which are undesirable from a fabrication perspective, and feature relatively small (${\lesssim}\,2\%$) bandgaps.
3D Chern insulators are characterized by a Chern vector $\mathbf{C} = (C_1, C_2, C_3)$ rather than a single Chern number, with components $C_i$ giving the Chern number of a $\kv$-slice orthogonal to $\Gv_i$~\cite{vanderbilt2018berry}.
Since the associated bands are assumed to be fully gapped, the slice-Chern numbers $C_i$ are invariant with respect to the slice position $k_i$.

Nontrivial Chern phases require breaking time-reversal symmetry.
To achieve this in the PhC context, we consider a gyroelectric material under a $\hat{\mathbf{z}}$-oriented external magnetic field with a permittivity tensor of the form~\cite{lu2013weyl}:
\begin{equation}
    \boldsymbol{\varepsilon}(g) = 
    \begin{bmatrix}
    \varepsilon_\perp & \iu g & 0 \\
    -\iu g & \varepsilon_\perp & 0 \\
    0 & 0 & \varepsilon_\parallel
    \end{bmatrix},
\end{equation}
where $\varepsilon_\perp = (\varepsilon_\parallel^2 + g^2)^{1/2}$.
We take $\varepsilon_\parallel=16$ and $g=12$ corresponding to a dimensionless effective magnetic field intensity of $|\mathbf{B}|={g}/{\varepsilon_\parallel}=0.75$.
Note that the determinant of $\boldsymbol{\varepsilon}(g)$ is independent of $g$ and $|\mathbf{B}|$, ensuring that the overall band structure is not shifted as a whole by the applied magnetic field~\cite{lu2013weyl}.

Although the slice-Chern numbers can be computed by discretization~\cite{fukui2015chern}, the computational cost is prohibitive for optimization purposes.
Instead, we focus our attention to PhCs in SGs 75 (P4) and 168 (P6), where the slice-Chern number $C_z$ can be computed from the bands' $n$-fold rotation-eigenvalues modulo $n$~\cite{fang2012bulk}, or equivalently, from the symmetry indicators $\nu_1^{(75)}\in \mathbb{Z}_4$ and $\nu_1^{(168)}\in\mathbb{Z}_6$.
Furthermore, under suitable convention choices, the symmetry indicators map directly to $C_z$, \ie $C_z^{(75)} = \nu_1^{(75)} \pmod 4$ and $C_z^{(168)} = \nu_1^{(168)} \pmod 6$~%
\footnote{%
    This mapping involves a fixed convention for the choice of nontrivial generators in \cref{eq:symmetryindicator}~\cite{song2018diagnosis}. 
    Specifically, we take
    $
    \mathbf{b}^{(75)}_1 = 
    - \hat{\mathbf{e}}_{\Gamma_2} + \hat{\mathbf{e}}_{\Gamma_4}
    - \hat{\mathbf{e}}_{Z_2} + \hat{\mathbf{e}}_{Z_4}
    $
    and
    $
    \mathbf{b}^{(168)}_1 = 
    + \hat{\mathbf{e}}_{H_2} - \hat{\mathbf{e}}_{H_3}
    + \hat{\mathbf{e}}_{K_2} - \hat{\mathbf{e}}_{K_3}
    - \hat{\mathbf{e}}_{L_1} + \hat{\mathbf{e}}_{L_2}
    - \hat{\mathbf{e}}_{M_1} + \hat{\mathbf{e}}_{M_2}
    $.
    Equivalently, the symmetry indicators are given by
    $
    \nu_1^{(75)} = 
    2n_{M_2} + n_{M_3} - n_{M_4} 
    - 2n_{\Gamma_1} - n_{\Gamma_3} + n_{\Gamma_4}
    - 2n_{X_1}
    \bmod 4
    $
    and
    $
    \nu_1^{(168)} = 
    n_{\Gamma_1} - 2n_{\Gamma_2} + 5n_{\Gamma_3} + 2n_{\Gamma_4} + 3n_{\Gamma_5}
    + 2n_{H_1} - 2n_{H_2}
    - 3n_{L_1}
    \bmod 6
    $, with irrep labels given in the CDML notation~\cite{cdml1979kroenecker}.
    }.
Lacking a corresponding $\kv$-slice with nontrivial rotation symmetry, the associated $C_{x,y}$ numbers are indeterminable from symmetry.

\Cref{fig:chern} summarizes the results of applying our technique to the optimization of 3D Chern insulating phases with target Chern vectors $\mathbf{C} = (0, 0, C_z)$ for $C_z = 1$ and $C_z = 2$.
For SG~75, we find PhC designs with $C_z = 1$ and a HS gap of $1.5\%$ between bands 4 and 5 [\cref{fig:chern}(c)].
Interestingly, the optimization finds a design with larger HS gap  of $2.1\%$ between bands 6 and 7 for a target Chern number of $C_z = 2$  [\cref{fig:chern}(d)].
Conversely, in SG~168, we are unable to find a $C_z=2$ design with a complete HS gap, but find a PhC design with a large $5.1\%$ $C_z = 1$ HS gap between bands 7 and 8 [\cref{fig:chern}(e)].

In more detail, we consider a $(010)$ interface with the applied magnetic field's sign flipped at the interface [\cref{fig:chern}(f)].
This flip preserves the band structure and, more specifically, the gap; but negates the associated Chern number, yielding a gap Chern number of $\Delta C_z = 1 - (-1) = 2$.
By the bulk--boundary correspondence principle, two chiral surface states must therefore traverse the gap.
To verify this, we computed the surface band structure across the remaining good wave numbers $(k_x,k_z)$ using a supercell of $8+8$ unit cells along $y$, as shown in \cref{fig:chern}(g).
A pair of chiral surface states traverse the gap, consistent with expectations, and effectively behave as a pair of 1D edge states at each $k_x$ value.
Additionally, we observe that the complete HS gap metric is highly effective in this example, in fact equaling the full BZ gap.
For computational reasons, our $y$-supercell was implemented as periodic; as a result, a set of redundant oppositely-traversing dispersion copies, residing at a second interface, were manually removed from the band structure visualization for clarity.

\section{Conclusion}

We have presented a combined global and local optimization framework which exploits several off-the-shelf optimization algorithms to effectively handle the non-continuous and non-differentiable objective functions that arise in the bandgap optimization of topological PhCs. 
We incorporate a level-set function expressed as a symmetry-constrained Fourier sum with a relatively low parameter-space dimensionality that makes optimization by gradient-free methods tractable, while simultaneously instating a symmetry-constrained search space. 
We have applied our method in several distinct symmetry settings, hosting several distinct types of nontrivial band topology.
In doing so, we have discovered several novel topological photonic designs featuring nodal lines, ideal Weyl points, and nontrivial 3D Chern insulators.
To our knowledge, these are the first proposed structures with nodal lines and Weyl points in the interior of the BZ, a design task that would be computationally intractable without the use of the symmetry-based topological band analysis tools that we use here.
As an example of the applicability and potential of our framework, our best Chern insulator design has the largest known bandgap of any 3D photonic Chern insulator, and achieves this without explicitly incorporating the supermodulation techniques employed in previous designs.

As we also discuss in the Results section, the HS bandgap maximization objective serves as an effective proxy for frequency-isolation of nodal lines and Weyl points. Nevertheless, this proxy can lead to non-ideal degeneracies that that overlap spectrally with other ``trivial'' bulk band features. 
Future work could focus on further optimization of the frequency isolation, especially in the local optimization stage given the computational cost of considering the full BZ. 
For example, an objective function for Weyl points could optimize the shape of the DOS directly such that it realizes the characteristic parabolic frequency-dependence over a desired range to exclude the possibility of Fermi pockets at or near the Weyl point frequency.
Analogously, a modified objective function for nodal lines could incorporate terms that minimize frequency dispersion across the line while simultaneously maximizing the bandgap to nearby Fermi pockets.

While we explore just three different kinds of band topology, the method we present is general and can be applied to any other topological feature that can be analyzed from band symmetry.
For example, the optimization scheme is not restricted to nodal lines and Weyl points in the interior of the BZ, but could be easily extended to nodal lines crossing or Weyl points lying on HS $\kv$-lines~\cite{zhang2020diagnosis}, simply by adjusting the objective function or symmetry setting accordingly.
More generally, other topological features of potential interest include the quantum spin Hall effect, higher-order topological insulators (\eg hinge states and corner states)~\cite{benalcazar2017electric,ota2019photonic}, and symmetry-protected degeneracies more broadly~\cite{yu2022encyclopedia}.
Furthermore, our exploration of optimization techniques can provide insight into bandgap optimization of non-topological (\ie trivial) 3D and 2D PhCs, as well as of other quasiparticles, such as in engineered phononic crystals.

The Fourier level-set parameterization and its incorporation of a maximal (spatial) frequency cutoff has the advantage that it generally avoids rapid structural variations and fine details.
Nevertheless, such constraints could potentially be incorporated with higher fidelity and stricter tolerances by exploiting standard techniques in topology optimization~\cite{sigmund2007morphology,lazarov2016length}.
In the context of PhCs, robust optimization techniques using min-max formulations for the objective function have been used to ensure satisfactory performance in the case of over- or underetching during fabrication~\cite{men2014fabrication,men2014robust}. 
In photonics design more broadly, density filters and projection steps can be applied periodically throughout optimization to remove small features and gaps~\cite{hammond2022high}.
A separate but related concern of particular importance for 3D structures is the incorporation of connectivity constraints to avoid floating structures or holes.
We note that several of our proposed Weyl point and Chern insulator structures exhibit floating features, reflecting the absence of a connectivity enforcement in the Fourier level-set parameterization.
While enforcement of connectivity constraints remains relatively underexplored in the 3D PhC context----perhaps in reflection of earlier topology optimizations of trivial bandgaps having automatically tended towards connected designs~\cite{men2014robust}---such constraints can in principle be explicitly incorporated~\cite{li2016structural}.
On the other hand, even PhCs with unconnected features may be realizable with emerging fabrication techniques such as implosion fabrication, wherein dielectric or metallic structures are embedded in a low-index hydrogel scaffold~\cite{oran20183d, mills2021implosion}.

In principle, any global or local optimization algorithm can be used in our framework.
Although we have explored several choices for each stage, we have not exhaustively compared different algorithms, and further improvements may well be achievable by simply using more optimal algorithms, \eg algorithms that explicitly handle non-continuous constraints.
Similarly, gradient-based optimization algorithms could be used in the local optimization stage to speed up convergence.
As a particularly exciting outlook, we note opportunities to exploit Bayesian optimization, a model-based algorithm typically used for global optimization of expensive non-convex functions. 
Recently, Bayesian optimization using Bayesian neural networks was demonstrated to outperform several gradient-free global optimization algorithms in the design of PhCs~\cite{kim2022deep}.
Bayesian optimization can also be extended to take advantage of gradient information~\cite{ament2022scalable} and non-continuous constraints~\cite{gelbart2015constrained} to improve convergence.

% ---------------------------------------
\section*{Acknowledgements}

The authors thank Charlotte Loh for helpful discussions.
This research is based upon work supported in part by 
the National Defense Science \& Engineering Graduate (NDSEG) Fellowship Program,
the Air Force Office of Scientific Research under the award number FA9550-20-1-0115 and FA9550-21-1-0299,
the US Office of Naval Research (ONR) Multidisciplinary University Research Initiative (MURI) grant N00014-20-1-2325 on Robust Photonic Materials with High-Order Topological Protection
and the U.S.\ Army Research Office through the Institute for Soldier Nanotechnologies at MIT under Collaborative Agreement Number W911NF-18-2-0048.
MIT SuperCloud and Lincoln Laboratory Supercomputing Center provided HPC resources that have contributed to the results reported in this paper.

% ---------------------------------------
\section*{Code availability}
The code underlying the results presented here will be made publicly available at \url{https://github.com/samuelkim314/topo-phc-opt}.

% ---------------------------------------
\bibliography{topo-opt}

\end{document}